\documentclass[amssymb,prl,twocolumn,floats]{revtex4}
\usepackage{bm}
\usepackage{graphicx}
\begin{document}

\title{Hall effect and magnetoresistance in p-type ferromagnetic
semiconductors}
\author{Tomasz Dietl}
\email{dietl@ifpan.edu.pl}
\homepage{http://www.ifpan.edu.pl/SL-2/sl23.html}
\affiliation{Institute of Physics, Polish Academy of Sciences,
\\ al. Lotnik\'{o}w 32/46, PL-02-668 Warszawa, Poland}
\author{Fumihiro Matsukura}
\affiliation{Institute of Electrical Communication, Tohoku
University, Katahira 2-1-1, Sendai 980-8577, Japan}
\affiliation{Institute of Physics, Polish Academy of Sciences, \\
al. Lotnik\'{o}w 32/46, PL-02-668 Warszawa, Poland}
\author{Hideo Ohno}
\affiliation{Institute of Electrical Communication, Tohoku
University, Katahira 2-1-1, Sendai 980-8577, Japan}
\author{ Jo\"el Cibert}
\affiliation{Laboratoire de Spectrom\'etrie Physique, CNRS -
Universit\'e Joseph-Fourier Grenoble Bo\^ite Postal 87, F-38402 St
Martin d'H\`eres Cedex, France}
\author{ David Ferrand}
\affiliation{Laboratoire de Spectrom\'etrie Physique, CNRS -
Universit\'e Joseph-Fourier Grenoble Bo\^ite Postal 87, F-38402 St
Martin d'H\`eres Cedex, France}


\begin{abstract}
Recent works aiming at understanding magnetotransport phenomena in
ferromagnetic III-V and II-VI semiconductors are described. Theory
of the anomalous Hall effect in p-type magnetic semiconductors is
discussed, and the relative role of side-jump and skew-scattering
mechanisms assessed for (Ga,Mn)As and (Zn,Mn)Te. It is emphasized
that magnetotransport studies of ferromagnetic semiconductors in
high magnetic fields make it possible to separate the contributions
of the ordinary and anomalous Hall effects, to evaluate the role of
the spins in carrier scattering and localization as well as to
determine the participation ratio of the ferromagnetic phase near
the metal-insulator transition. A sizable negative
magnetoresistance in the regime of strong magnetic fields is
assigned to the weak localization effect.
\end{abstract}
\maketitle

\section{Introduction}

The assessment of magnetic characteristics by means of
magnetotransport studies is of particular importance in the case of
thin films of diluted magnets, in which the magnitude of the total
magnetic moment is typically small. For this reason, recent years
have witnessed a renewed interest in the nature of the anomalous
Hall effect (AHE) \cite{Hirs99,Ye99,Zhan00,Crup01,Jung02},
which--if understood theoretically--can serve to determine the
magnitude of magnetization. Also magnetoresistance, though less
directly, provides information on the magnetism and on the
interplay between electronic and magnetic degrees of freedom.

In this paper, we discuss selected magnetotransport properties of
III-V and II-VI magnetic semiconductors containing Mn as the
magnetic element.  In particular, we show that the side-jump
mechanism  accounts for the  magnitude of the anomalous Hall effect
in both (Ga,Mn)As and (Zn,Mn)Te samples for which extensive
experimental data are available. We emphasize, however, that the
current theory of the effect requires further refinements. We also
suggest that weak localization magnetoresistance may contribute to
the increase of the hole conductivity in the limit of low
temperatures $T$ and high magnetic fields $H$. Recent review papers
\cite{Ohno01,Diet02} summarize rather thoroughly principal findings
of previous comprehensive studies of these materials, which are not
touched upon here.

\section{Hall effect in ferromagnetic semiconductors -- theoretical models}

The Hall resistance $R_{Hall}\equiv \rho_{yx}/d$ of a film of the
thickness $d$ is empirically known to be a sum of ordinary and
anomalous Hall terms in magnetic materials \cite{Chie80},
\begin{equation}
R_{Hall} = R_0\mu_oH/d + R_S\mu_oM/d.
\end{equation}
Here, $R_0$ and $R_S$ are the ordinary and anomalous Hall
coefficients, respectively ($R_0> 0$ for the holes), and $M(T,H)$
is the component of the magnetization vector perpendicular to the
sample surface. While the ordinary Hall effect serves to determine
the carrier density, the anomalous Hall effect (known also as the
extraordinary or spin Hall effect) provides valuable information on
magnetic properties of thin films. The coefficient $R_S$ is usually
assumed to be proportional to $R_{sheet}^{\alpha}$, where
$R_{sheet}(T,H)$ is the sheet resistance and the exponent $\alpha$
depends on the mechanisms accounting for the AHE.

If the demagnetization effect were been dominating, $R_S$ would be
rather proportional to $R_0$ than to $R_{sheet}$. However, there is
no demagnetization effect in the magnetic field perpendicular to
the film surface, $B=\mu_oH$. Here, spin-orbit interactions control
totally $R_S$. In such a situation $\alpha$ is either 1 or 2
depending on the origin of the effect: the skew-scattering
mechanism, for which the Hall conductivity is proportional to
momentum relaxation time $\tau$, results in $\alpha \approx 1$
\cite{Chie80,Lutt58,Lero72,Nozi73,Chaz75}. From the theory point of
view particularly interesting is the side-jump mechanism. This is
because in both weak and strong scattering limit, $\omega \tau \gg
1$ and $\omega \tau \ll 1$, where $ \omega$ is the frequency of the
electric field, the corresponding Hall conductivity
$\sigma_{AH}=R_SM/(R_{sheet}d)^2]$ does not depend explicitly on
scattering efficiency but only on the band structure parameters
\cite{Lutt58,Nozi73,Chaz75}. Surprisingly, $\sigma_{AH}(\omega \tau
\gg 1) =- \sigma_{AH}(\omega \tau \ll 1)$ according to these works.

For both skew-scattering and side-jump mechanisms, the overall
magnitude of the anomalous Hall resistance depends on the strength
of the spin-orbit interaction and spin polarization of the carriers
at the Fermi surface. Accordingly, at given magnetization $M$, the
effect is expected to be much stronger for the holes than for the
electrons in tetrahedrally coordinated semiconductors. For the
carrier-mediated ferromagnetism, the latter is proportional to the
exchange coupling of the carriers to the spins, and varies -- not
necessarily linearly -- with the magnitude of spin magnetization
$M$. Additionally, the skew-scattering contribution depends on the
asymmetry of scattering rates for particular spin subbands, an
effect which can depend on $M$ in a highly nontrivial way.
Importantly, the sign of either of the two contributions can be
positive or negative depending on a subtle interplay between the
orientations of orbital and spin momenta as well as on the
character (repulsive vs. attractive) of scattering potentials.

We presume that general theory of the AHE effect in semiconductors
\cite{Nozi73,Chaz75} gives correctly the ratio of side-jump and
skew-scattering mechanisms, also in the case of p-type
semiconductors. If scattering  by ionized impurities dominates,
this ratio is then given by \cite{Lero72,Chaz75,Chaz74},
\begin{equation}
\frac{\sigma_{AH}^{sj}}{\sigma_{AH}^{ss}}= \pm f(\xi)(N_A+N_D)
/(pr_sk_F\ell),
\end{equation}
where the positive sign corresponds to the weak scattering limit.
Here, $f(\xi) \approx 10$ is a function that depends weakly on the
screening dimensionless parameter $\xi$; $(N_A+N_D)/p$ is the ratio
of the ionized impurity and carrier concentrations; $r_s$ is the
average distance between the carriers in the units of the effective
Bohr radius, and $\ell$ is the mean free path. Similarly, for
spin-independent scattering by short range potentials, $V({\bm
{r}}) = V\delta(\bm{r} - \bm{r}_i)$ \cite{Nozi73},
\begin{equation}
\frac{\sigma_{AH}^{sj}}{\sigma_{AH}^{ss}}= \pm 3/[\pi
V\rho(\varepsilon_F)k_F\ell],
\end{equation}
where the negative sign corresponds to the weak scattering limit
and $\rho(\varepsilon_F)$ is the density of states at the Fermi
level. Of course, the overall sign depends on the sign of the
scattering potential $V$.

In order to find out which of the two AHE mechanisms operates
predominantly in p-type tetrahedrally coordinated ferromagnetic
semiconductors, we note that scattering by ionized impurities
appears to dominate in these heavily doped and compensated
materials. This scattering mechanism, together with alloy and spin
disorder scattering, limits presumably the hole mobility and leads
ultimately to the metal-to-insulator transition (MIT). Since at the
MIT $r_s\approx 2$ and $k_F\ell \approx 1$ we expect from Eq.~2
that as long as the holes remain close to the localization boundary
the side-jump mechanism accounts for the AHE. It would be
interesting on know how quantum localization corrections affect the
anomalous Hall conductivity as well as how to extend theory towards
the insulator side of the MIT. A work in this direction has
recently been reported \cite{Duga01}.

Recently, Jungwirth {\it et al.} \cite{Jung02} developed a theory
of the AHE in p-type zinc-blende magnetic semiconductors, and
presented numerical results for the case of (Ga,Mn)As, (In,Mn)As,
and (Al,Mn)As. The employed formula for $\sigma_{AH}$ corresponds
to that given earlier \cite{Lutt58,Nozi73,Chaz75} for the side-jump
mechanism in the weak scattering limit. For the hole concentration
$p$ such that the Fermi energy is much smaller than the spin-orbit
splitting $\Delta_o$ but larger than the exchange splitting $h$
between the majority $j_z=-3/2$ and minority $j_z=+3/2$ bands at
$k=0$, $\Delta_o \gg |\epsilon_F| \gg h$, Jungwirth {\it {\it et
al.}} \cite{Jung02} predict within the $4\times 4$ spherical
Luttinger model
\begin{equation}
\sigma_{AH}^{sj}=e^2hm_{hh}/[4\pi^2\hbar^3(3\pi p)^{1/3}].
\end{equation}
Here the heavy hole mass $m_{hh}$ is assumed to be much larger than
the light hole mass $m_{lh}$, whereas $\sigma_{AH}^{sj}$ becomes by
the factor of $2^{4/3}$ greater in the opposite limit
$m_{hh}=m_{lh}$. In the range $h \ll |\epsilon_F| \ll \Delta_o$ the
determined value of $\sigma_{AH}^{sj}$ is positive, that is the
coefficients of the normal and anomalous Hall effects are expected
to have the same sign. However, if the Fermi level were approached
the split-off $\Gamma_7$ band, a change of sign would occur.

We have derived $\sigma_{AH}^{sj}$ from Chazalviel's formula
\cite{Chaz75} in the weak scattering limit (which is equivalent to
Eq.~4 of Jungwirth {\it et al.} \cite{Jung02}), employing the known
form of the heavy hole Bloch wave functions $u_{{\bm k},j_z}$
\cite{Szym78}.  Neglecting a small effect of the spin splitting on
the heavy hole wave functions, we find $\sigma_{AH}^{sj}$ to be
given by the right hand side of Eq.~4 multiplied by the factor
$(16/9)\ln2-1/6\approx 1.066$.

Obviously, the presence of the AHE makes a meaningful determination
of the carrier type and density difficult in ferromagnetic
semiconductors. Usually, the ordinary Hall effect dominates only in
rather high magnetic fields or at temperatures several times larger
than $T_C$. It appears, therefore, that a careful experimental and
theoretical examination of the resistivity tensor in wide field and
temperature ranges is necessary to separate characteristics of the
spin and carrier subsystems.

\section{Comparison between theoretical and experimental results: (Ga,Mn)As}

As mentioned above, because of the dominance of the anomalous Hall
term in wide temperature and field ranges, it is not
straightforward to determine the carrier type and concentration in
ferromagnetic semiconductors. Only at low temperatures and under
very high fields, the anomalous Hall term saturates, so that the
ordinary Hall coefficient can be determined from the remaining
linear change of the Hall resistance in the magnetic field. Note
that although magnetization saturates in relatively low magnetic
fields, the negative MR usually persists, and generates the field
dependence of the anomalous Hall coefficient.

\begin{figure}
\includegraphics*[width=75mm]{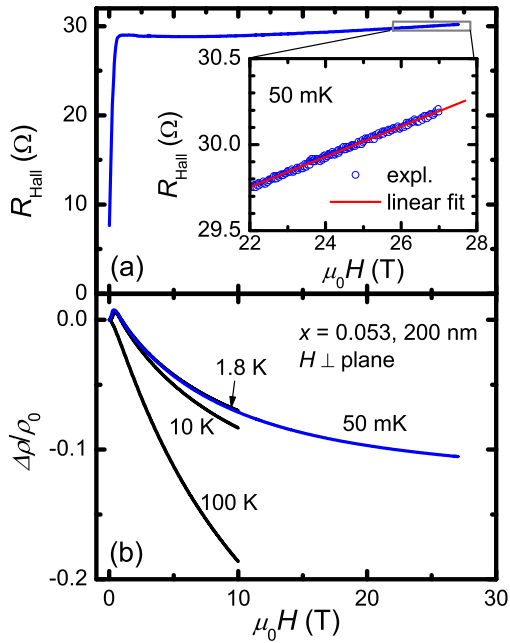}
\caption[]{Magnetotransport properties of 200-nm thick film of
Ga$_{1-x}$Mn$_x$As with $x = 0.053$ at 50~mK in high magnetic
fields. (a) Hall resistance, which is a linear function of the
magnetic field in the high-field region (inset). (b) Sheet
resistance (after \cite{Omiy00}).}
\end{figure}

Measurements of $R_{Hall}$ at 50~mK in the field range of 22--27~T
on the sample with $x = 0.053$ revealed that the conduction is
p-type, consistent with the acceptor character of Mn, as shown in
Fig.~1 \cite{Omiy00}. The determined hole concentration is $p =
3.5\times 10^{20}$ cm$^{-3}$, about 30\% of the Mn concentration. A
similar value of the hole concentration, which is almost
independent of $x$, has been obtained from the Seebeck coefficient
assuming a simple model of the valence band \cite{Osin01}. If all
Mn centers are acting as acceptors in the metallic sample described
above, 70\% of them must have been compensated by donors. The most
natural candidates for these donors are As antisite defects, which
act as deep donors in GaAs. Accordingly, (Ga,Mn)As should become
insulating at room temperature when the density of As antisites
exceeds the density of shallow acceptors. Because the magnitudes of
these densities are comparable and moreover fluctuate from run to
run depending on subtleties of the growth conditions, we expect the
overcompensation to occur occasionally. However, no such
'overcompensated' sample has been obtained so far. This seems to
call for mechanisms controlling the upper limit of the excess As
concentration and/or leading to selfcompensation of Mn but not to
overcompensation. One candidate for the latter might be the Mn
interstitial, which acts as the relevant compensating donor
according to first principles calculations \cite{Mase01} and recent
channeling studies \cite{Yu02}.

\begin{figure}
\includegraphics*[width=80mm]{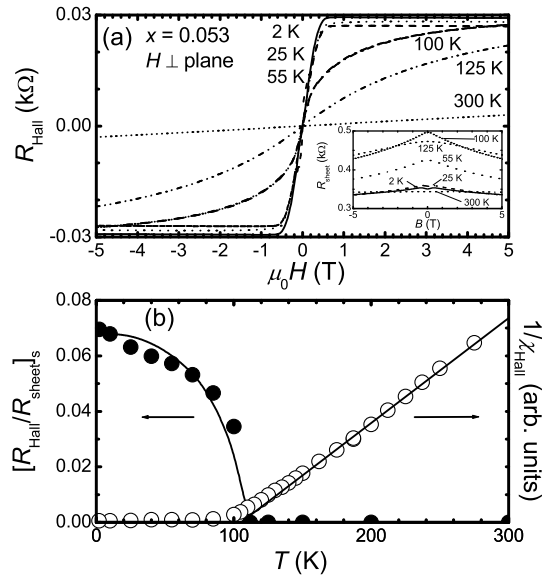}
\caption[]{Temperature dependence of the Hall resistance $R_{Hall}$
for the same sample as in Fig.~1. The inset shows the temperature
dependence of the sheet resistance $R_{sheet}$. (b) Temperature
dependence of the saturation magnetization $R_{Hall} /R_{sheet}$
obtained by using Arrott plots (closed circles) and inverse
susceptibility $1/R_{Hall}$ (open circles), both from the transport
data shown in (a). Solid lines depict [$R_{Hall}/R_{sheet}$] and
$1/R_{Hall}$ calculated assuming the mean-field Brillouin behavior
for the Mn spin $S = 5/2$ and the Curie-Weiss law, respectively
(after \cite{Mats98}).}
\end{figure}

Figures 2 and 3 present a comparison of the Hall resistance
$R_{Hall}$ \cite{Mats98} and magnetization $M$ from SQUID
measurements \cite{Ohno01} at various temperatures plotted as a
function of the magnetic field for the same 200-nm thick
Ga$_{0.947}$Mn$_{0.053}$As film. The inset shows the temperature
dependence of $R_{sheet}$. A general similarity between
$R_{Hall}(T,H)$ and $M(T,H)$ confirms that the contribution from
the ordinary Hall term is rather small in the displayed field and
temperature range. If $R_{sheet}$ depends on temperature, a
comparison of magnetization and magnetotransport data can serve to
identify whether the skew-scattering or side-jump mechanism
dominates. In particular, since $R_{Hall}/R_{sheet}^{\alpha} \sim
M$, Arrott's plots can be employed to determine the temperature
dependence of spontaneous magnetization $M_S(T) = M(T,0)$.  As
shown in Fig.~2, the temperature dependence of $M_S$ determined by
the magnetotransport measurements assuming $\alpha = 1$ can be
fitted rather well by the mean-field Brillouin function
\cite{Mats98}.  A different temperature dependence stems from
direct magnetization measurements in a SQUID magnetometer presented
in Fig.~3 for the same sample \cite{Ohno01}. Owing to an increase
of $R_{sheet}$ with temperature in this sample, $M_S(T)$ determined
by the two methods can be made somewhat closer by choosing $\alpha
= 2$. This may indicate that the side-jump mechanism dominates. The
dependence $M_S(T)$ determined by the SQUID measurements cannot be
fitted by a simple Brillouin function, $M_S(T)/M_S(0) = 1 -
(T/T_C)^{\gamma}$, where $\gamma =5/2$. Actually, a less convex
dependence, $n<5/2$, is expected even within the MFA in magnetic
semiconductors \cite{Diet01}.

\begin{figure}
\includegraphics*[width=90mm]{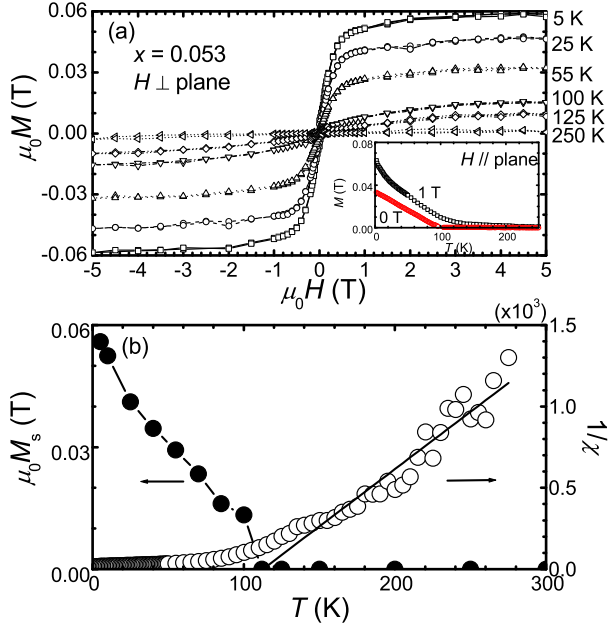}
\caption[]{Temperature dependence of magnetization for 200-nm thick
Ga$_{1-x}$Mn$_x$As with $x$ = 0.053. Magnetic field is applied
perpendicular to the sample surface (hard axis). Inset shows
temperature dependence of remanent magnetization (0~T) and
magnetization at 1~T in the field parallel to the film surface. (b)
Temperature dependence of saturation magnetization $M_S$ determined
from the data shown in (a) by using the Arrott plots (closed
circles). Open circles show inverse magnetic susceptibility and the
Curie-Weiss fit is depicted by solid straight line (after
\cite{Ohno01}).}
\end{figure}

The findings presented above have been exploited by Jungwirth et
al. \cite{Jung02} to test their theory of the AHE. The results of
such a comparison are shown in Fig.~4 \cite{Jung02}. There is a
good agreement between the theoretical and experimental magnitude
of the Hall conductivity. Importantly, no significant contribution
from the skew scattering is expected for the (Ga,Mn)As sample in
question, for which, according to Figs.~1-3, $(N_A+N_D)/p \approx
5$, $r_s \approx 1.1$, and $k_F\ell \approx 0.8$, so that
$\sigma_{AH}^{sj}/\sigma_{AH}^{ss}\approx 57$. Finally, we note
that the sign of the effect indicates that weak scattering limit
$\omega \tau \gg 1$ is appropriate in the case under consideration.
Obviously, however, further works are necessary to elucidate the
role of intra- and inter-subband scattering processes in the
physics of the side-jump mechanism.

\begin{figure}
\includegraphics*[width=120mm]{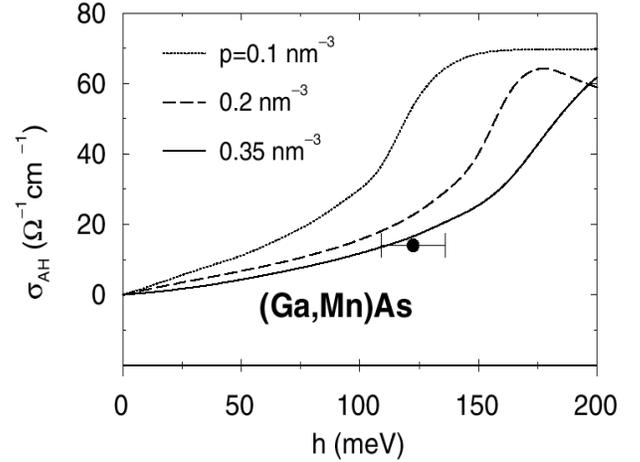}
\caption[]{Full numerical simulations of the anomalous hall
conductivity $\sigma_{AH}$ for GaAs host with hole densities $p =
10^{20}$, (dotted lines), $2\times 10^{20}$ (dashed lines), and
$3.5\times 10^{20}$ cm$^{-3}$ (solid lines).  Filled circle
represents measured Hall conductivity (Fig.~2). The saturation
mean-field value of the splitting $h$ between $\Gamma_8$ heavy hole
subbands  was estimated from nominal sample parameters. Horizontal
error bar corresponds to the experimental uncertainty of the p$-$d
exchange integral. Experimental hole density in the (Ga,Mn)As
sample is $3.5\times 10^{20}$ cm$^{-3}$ (after \cite{Jung02}).}
\end{figure}

It is important to note that there exist several reasons causing
that the Hall effect and direct magnetometry can provide different
information on magnetization. Indeed, contrary to the standard
magnetometry, the AHE does not provide information about the
magnetization of the whole samples but only about its value in
regions visited by the carriers. Near the metal-insulator boundary,
especially when the compensation is appreciable, the carrier
distribution is highly non-uniform. In the regions visited by the
carriers the ferromagnetic interactions are strong, whereas the
remaining regions may remain paramagnetic. Under such conditions,
magnetotransport and direct magnetic measurements will provide
different magnetization values \cite{Diet00}. In particular, $M_S$
at $T \rightarrow 0$, as seen by a direct magnetometry, can be much
lower than that expected for a given value of the magnetic ion
concentration. High magnetic fields are then necessary to magnetize
all localized spins. The corresponding field magnitude is expected
to grow with the temperature and strength of antiferromagnetic
interactions that dominate in the absence of the holes.

Finally, we note that no clear indication of the presence of MnAs
clusters has been observed in the transport studies, even in the
cases, where direct magnetization measurements detect their
presence. One of possibilities is that the Schottky barrier
formation around the MnAs clusters prevents their interaction with
the carriers. Conversely, the presence of a clear influence of the
magnetic subsystem onto transport properties (colossal
magnetoresistance, anomalous Hall effect) can be taken as an
evidence for the mutual interactions of the spins and the carriers.
Such interactions are behind virtually all proposed applications of
magnetic semiconductors.

\section{Experimental results: (Zn,Mn)Te}

Figure 5 shows the Hall resistivity $R_{Hall}$ measured at various
temperatures for the highly doped Zn$_{0.981}$Mn$_{0.019}$Te:N
sample \cite{Ferr00}. The quoted hole concentration is deduced from
the slope of the room temperature Hall resistance. The dependence
$R_{Hall}$ is linear in the magnetic field and temperature
independent down to 150~K. In the case of the p-ZnTe sample, this
normal Hall effect $R_{Hall}$, linear in the field $H$ and
temperature independent, is observed down to 1.6~K. By contrast, in
the case of p-Zn$_{1-x}$Mn$_{x}$Te, when decreasing the temperature
below 100~K, one observes first an increase of the slope of the
Hall resistance, and then a strong non-linearity, which point to
the presence of the anomalous Hall effect. As expected, no
anomalous Hall effect has been detected in wide-gap n-type II-VI
DMS \cite{Shap86}. At low temperature and high field, the Mn or the
hole spin polarization saturate, and then the Hall resistivity
exhibits again a linear dependence on the applied field, with the
same slope as at room temperature. Thus, while the spin-dependent
component is too large to allow us to determine the hole density at
low temperatures and in small fields, due to low $T_C$, its
magnitude becomes negligibly small at room temperature, or at
low-temperature in high fields. For these two cases, the slope of
the Hall resistance was found to be identical, giving unambiguously
the value of the hole density.

In the case of less doped samples, it was possible to measure the
Hall resistivity down to typically 10~K, with the same conclusions,
{\it i.e.}, (i) the normal Hall effect dominates at temperatures
above 150~K; (ii) the Hall resistivity varies linearly with the
magnetic field at low temperature in sufficiently large magnetic
fields, and (iii) a strong spin-dependent component appears at weak
magnetic fields and at low temperatures, though its accurate
determination in this region is hampered by the large value of the
resistance and a strong magnetoresistance. As mentioned above, the
Hall resistance provides direct information on the degree of spin
polarization $\cal{P}$ of the carrier liquid.

\begin{figure}
\includegraphics*[width=85mm]{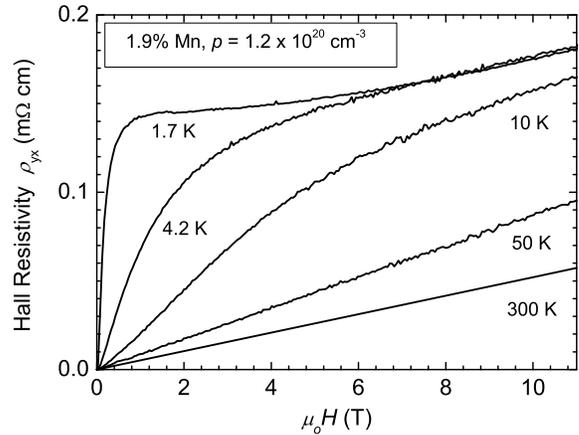}
\caption[]{Hall resistivity versus magnetic field at different
temperatures, from room temperature down to 1.7~K in metallic
p-Zn$_{0.981}$Mn$_{0.019}$Te:N. The nonlinear temperature dependent
component is assigned to the anomalous Hall effect, which strongly
increases on approaching the ferromagnetic phase transition (after
\cite{Ferr01}).}
\end{figure}

In Fig. 6, $\rho_{yx}/\rho_{xx} - \mu B$, i.e., the spin dependent
Hall angle, is compared to the magnetization measured in a
vibrating sample magnetometer \cite{Ferr00}. The normal Hall angle
$\mu B =\mu\mu_oH$ was subtracted assuming a constant hole mobility
$\mu$~i.e., assigning the conductivity changes entirely to
variations in the hole concentration. This assumption is not
crucial for the present highly doped sample, but it proves to be
less satisfactory for the less doped samples. As shown in Fig.~6, a
reasonable agreement is found by taking,
\begin{equation}
\rho_{yx}/\rho_{xx} = \mu B + \Theta M/M_S,
\end{equation}
where $M_S$ is the saturation value of magnetization and $\Theta =
0.04$ is the adjustable parameter.  For the sample in question, the
maximum value of hole polarization, $(p^{up}-p^{down})/(p^{up}
+p^{down})$, has been estimated to be of the order of 10\%
\cite{Ferr00}.

\begin{figure}
\includegraphics*[width=90mm]{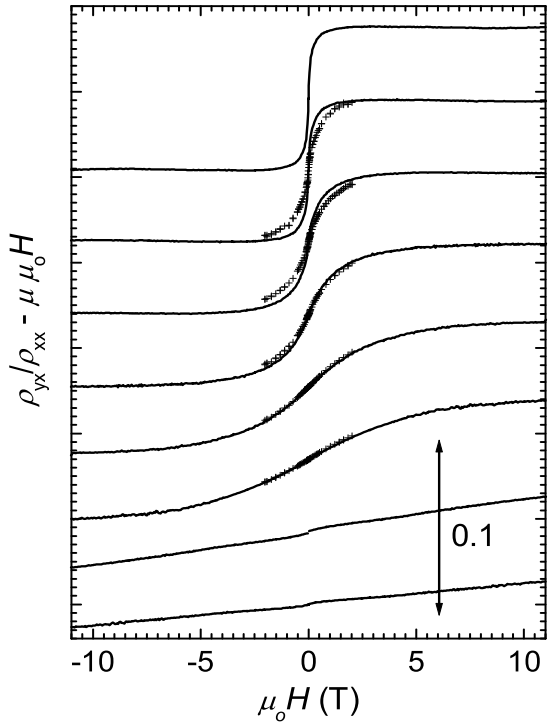}
\caption[]{Comparison of the normalized
anomalous Hall effect (lines) with the normalized magnetization
$M/M_S$ (crosses); from top to bottom: 1.7, 2.8, 4.2, 7, 10, 30,
and 50~K; the data are shifted for clarity (after \cite{Ferr00}).}
\end{figure}

We note that similarly to the case of (Ga,Mn)As, the sign and
magnitude of the anomalous Hall coefficient suggests that the side
jump mechanism in the weak scattering limit is involved. We
evaluate $\Theta$ theoretically from Eq.~4 by adopting parameters
suitable for the sample in question, $m_{hh} = 0.6m_o$, $\rho_{xx}
= 5\times 10^{-3}$~$\Omega$cm and the saturation value of the
splitting $h=41$ meV. This leads to $\sigma_{AH}^{sj}
=13.1$~($\Omega$cm)$^{-1}$ and $\Theta^{sj} = 0.065$, in a
reasonable agreement with the experimental value $\Theta = 0.04$.
Since a contribution from the light hole band will enhance the
theoretical value, we conclude that the present theory describes
the anomalous hole effect within the factor of about two. We note
also that in contrast to earlier suggestions \cite{Ferr00}, not
skew-scattering but the side-jump mechanism appears to give the
dominant contribution to the AHE in p-(Zn,Mn)Te. However, as
mentioned above, further theoretical work is needed to assess the
role of hole scattering.

\section{Magnetoresistance}

There is a number of effects that can produce a sizable
magnetoresistance in magnetic semiconductors, especially at the
localization boundary \cite{Diet94}. In particular, spin disorder
scattering shifts the MIT towards higher carrier concentration.
Since the magnetic field orders the spins, negative
magnetoresistance occurs, sometimes leading to the field-induced
insulator-to-metal transition \cite{Kats98,Ferr01}. Deeply in the
metallic phase, virtually all spins contribute to the ferromagnetic
ordering. Critical scattering and the associated negative
magnetoresistance are then observed \cite{Omiy00}. However, as
shown in Fig.~1, the negative magnetoresistance hardly saturates,
even in the extremely strong magnetic fields. In order to explain
this observation we note that the giant splitting of the valence
band makes both spin-disorder and spin-orbit scattering relatively
inefficient. Under such conditions, weak localization
magnetoresistance can show up at low temperatures, where inelastic
scattering ceases to operate.  According to Kawabata \cite{Kawa80},
\begin{equation}
\Delta \rho/\rho = -n_ve^2C_o\rho(e B/\hbar)^{1/2}/(2\pi^2\hbar),
\end{equation}
where $C_o \approx 0.605$ and $1/2 \leq n_v \leq 2$ depending on
whether one or all four hole subbands contribute to the charge
transport. For the sample in question the above formula gives
$\Delta \rho/\rho = - 0.1$ for $n_v = 1$ and 25~T, the value
consistent with the experimental results in Fig.~1. Since the
negative magnetoresistance takes over above $B_i \approx 1$~T, we
can evaluate a lower limit for the spin-flip scattering time
\cite{Kawa80,Ono84,Sawi86}, $\tau_{s} >  m*/(eB_i k_F\ell) \approx
5$~ps for $m*=0.7m_o$ and $k_F\ell = 0.8$.

\section{Summary}

Experimental results discussed above demonstrate the critical
importance of the Hall effect in the assessment of the magnetic
properties of III-V ferromagnetic semiconductors. Furthermore, they
suggest that the side-jump mechanisms gives the dominant
contribution for metallic samples, in which a comparison between
theoretical expectations and experimental results is possible.
Importantly, the theory discussed here explains the sign of the
effect and, together with the results obtained by Jungwirth {\it et
al.}~\cite{Jung02}, explains the magnitude of the Hall conductance.

Importantly, such studies can also serve to detect a participation
of the double exchange mechanism in the spin-spin interactions.
This is because, the spin excitations associated with this coupling
produce a strong temperature dependence of $R_S$ near $T_C$
\cite{Ye99}.  We take the absence of a strong temperature
dependence of $R_S$ near $T_C$ as an evidence for the minor
importance of the double exchange in the studied systems.
Conversely,  a good agreement between the measured and calculated
Hall coefficients,  if confirmed by further investigations, will
constitute an important support for basic assumptions behind the
Zener model \cite{Diet00} of ferromagnetism in this class of
ferromagnetic semiconductors.

Furthermore, the accumulated information on magnetoresistance
points to significance of the spin-disorder scattering as well as
reveal various effects associated with the interplay between spin
and localization phenomena, specific to doped diluted magnetic
semiconductors in the vicinity of the metal-insulator transition.

\section*{Acknowledgments}

We thank Tomas Jungwirth, Allan H. MacDonald, and Jairo Sinova for
valuable discussions on the physics of the anaomalous Hall effect.
The work was supported by Foundation for Polish Science, by State
Committee for Scientific Research, Grant No.~2-P03B-02417 as well
as by FENIKS project (EC: G5RD-CT-2001-00535).

\end{document}